\documentclass[aps,twocolumn,showpacs,preprintnumbers]{revtex4}

\usepackage[utf8]{inputenc}
\usepackage[english]{babel}
\usepackage{amsmath}
\usepackage{braket}
\usepackage{bm}
\usepackage{chemformula}
\usepackage{siunitx}
\usepackage{amsbsy}

\usepackage{graphicx}
\usepackage{multirow}
\usepackage{fixltx2e}
\usepackage{booktabs}
\usepackage[T1]{fontenc}
\usepackage{dcolumn}  
\usepackage{float}

\usepackage{amssymb}
\usepackage{amsfonts,dsfont}
\usepackage{subfigure}

\usepackage[normalem]{ulem}
\usepackage[bbgreekl]{mathbbol}
\usepackage{mathrsfs,empheq}
\usepackage{relsize,scalerel}

\newcommand{\eq}{\sss{eq}}

\newcommand{\pstmmc}{\textit{P}6\textsubscript{3}\textit{/mmc}}
\newcommand{\ScH}{ScH\textsubscript{6}}

\newcommand{\Rcal}{\mathcal{R}}

\newcommand{\Fcal}{\mathcal{F}}

\newcommand{\bR}{\boldsymbol{R}}

\newcommand{\bD}{\boldsymbol{D}}

\newcommand{\bq}{\boldsymbol{q}}
\newcommand{\bk}{\boldsymbol{k}}

\newcommand{\bvarPhi}{\boldsymbol{\varPhi}}

\newcommand{\bRcal}{\boldsymbol{\Rcal}}

\newcommand{\sss}[1]{\scriptscriptstyle{\text{#1}}}

\newcommand{\rscha}{\bRcal}
\newcommand{\rschatrial}{\bRcal}
\newcommand{\phischatrial}{\bvarPhi}

\newcommand{\rhoschatrial}{{\tilde\rho}_{\scriptscriptstyle{\rscha},\scriptscriptstyle{\phischatrial}}}

\newcommand{\bRcaleq}{\bRcal_{\eq}}
\newcommand{\DF}{D^{\sss{(F)}}}
\newcommand{\DHA}{D^{\sss{(H)}}}
\newcommand{\bDF}{\bD^{\sss{(F)}}}

\begin{document}

\title{Quantum anharmonic enhancement of superconductivity in \pstmmc\ \ScH\ at high pressures: a first-principles study}

\author{Pugeng Hou$^{1}$, Francesco Belli$^{2,3}$, Raffaello Bianco$^{3}$, Ion Errea$^{* 2,3,4}$}

\affiliation {\it 1 College of Science, Northeast Electric Power University, Changchun Road 169, 132012, Jilin, P. R. China} 

\affiliation {\it 2 Fisika Aplikatua Saila, Gipuzkoako Ingeniaritza Eskola, University of the Basque Country (UPV/EHU), Europa Plaza 1, 20018 Donostia/San Sebastián, Spain} 

\affiliation {\it 3 Centro de Física de Materiales (CSIC-UPV/EHU), Manuel de Lardizabal Pasealekua 5, 20018 Donostia/San Sebastián, Spain} 

\affiliation {\it 4 Donostia International Physics Center (DIPC), Manuel de Lardizabal Pasealekua 4, 20018 Donostia/San Sebastián, Spain}

\date{\today}

\begin{abstract}  
Making use of first-principles calculations, we analyze the effect of quantum ionic fluctuations and lattice anharmonicity on the crystal structure and superconductivity of \pstmmc\ \ScH\ in the 100–160 GPa pressure range within the stochastic self-consistent harmonic approximation. We predict a strong correction to the crystal structure, the phonon spectra, and the superconducting critical temperatures, which have been estimated in previous calculations without considering ionic fluctuations on the crystal structure and assuming the harmonic approximation for the lattice dynamics. Quantum ionic fluctuations have a large impact on the H$_2$ molecular-like units present in the crystal by increasing the hydrogen-hydrogen distance about a 5\%. According to our anharmonic phonon spectra, this structure will be dynamically stable at least above 85 GPa, which is 45 GPa lower than the pressure given by the harmonic approximation. Contrary to many superconducting hydrogen-rich compounds, where quantum ionic effects and the consequent anharmonicity tend to lower the superconducting critical temperature, our results show that it can be enhanced in \pstmmc\ \ScH\ by approximately a 15\%. We attribute the enhancement of the critical temperature to the stretching of the H$_2$ molecular-like units and the associated increase of the electron-phonon interaction. Our results suggest that quantum ionic effects increase the superconducting critical temperature in hydrogen-rich materials with H$_2$ units by increasing the hydrogen-hydrogen distance and, consequently, the electron-phonon interaction.
\end{abstract}

\maketitle 

\section{Introduction}

The study of hydrogen-rich compounds at high pressure is mostly motivated by their potential to be high-temperature superconductors. In 2004, Ashcroft suggested that hydrogen-rich metallic alloys may become metallic and superconducting at lower pressures than pure hydrogen\cite{1}. A number of hydrogen-rich compounds were subsequently predicted by {\it ab initio} calculations to be good superconductors with critical temperatures ($T_c$) reaching remarkable high values \cite{2,3,4,5,6,7,8,9,10,11,12,13,14,15,16,17,18,19,20,21}. Experimentally high critical temperatures have been reported and observed above 200 K in the last years, for instance, in sulfur\cite{22}, lanthanum\cite{23,24}, and yttrium\cite{25,26,27} superhydrides at pressures  exceeding 100 GPa. More recently, a ternary compound based on C, H, and S has finally reached room-temperature superconductivity at pressures above 250 GPa\cite{28}, breaking the highest $T_c$ record. 

The potential of first-principles calculations based on density-functional theory (DFT) to guide the experimental work on the right track is clear now\cite{19,20,21}. Indeed,  many of the experimental discoveries had been anticipated by DFT calculations\cite{8,9,12,14,15}. This means that having accurate methods for predicting the structural and superconducting properties of hydrogen-rich compounds is essential for the advent of new discoveries, hopefully, at lower pressures. Most of the structural and superconducting predictions done so far rely on the classical picture in which the atoms lay at the $\bR_0$ minimum of the Born-Oppenheimer potential $V(\bR)$, where $\bR$ represents the position of all atoms, and oscillate around these equilibrium positions with phonon frequencies determined from the $\left[ \frac{\partial^2V(\bR)}{\partial R^a \partial R^b} \right]_{\bR_0}$ force constants (harmonic approximation). Let $a$ and $b$ be indexes that label both an atom in the crystal and a Cartesian direction. However, it has been shown recently that quantum ionic fluctuations and the consequent anharmonicity can have a huge impact both on the crystal structure and the phonon spectrum of hydrogen-rich compounds, strongly affecting the predicted $T_c$\cite{44,34,35,38,39,41,25,monacelli2019black,hou2021strong}, due to the lightness of hydrogen. Interestingly, these effects can both strongly enhance or suppress the predicted critical temperature. In aluminium\cite{hou2021strong,33}, palladium\cite{34,35}, and platinum\cite{35} hydrides anharmonicity hardens the H-character optical modes and suppresses superconductivity in a large degree. In the possible metallic and molecular $Cmca$ phase of hydrogen the situation is the contrary\cite{41}: anharmonicity doubles $T_c$ bringing it from around 100 K to values well above 200 K. Also in LaH$_{10}$, it has been argued that quantum effects stabilize a crystal structure with huge electron-phonon coupling that otherwise would be unstable. How quantum anharmonic effects affect superconductivity is thus not known {\it a priori}.

In this work we present a first-principles analysis of the role of quantum ionic fluctuations and anharmonicity on the recently predicted high-temperature superconducting \ScH\ in the \pstmmc\ phase at high pressures\cite{29,30,31}. This metallic compound, which contains H$_2$-like molecular units, has been proposed to be thermodynamically and dynamically stable above 135 GPa with an estimated $T_c$ of 63 K at 145 GPa\cite{29}. Other theoretical calculations also within the classical harmonic approach\cite{30} suggest that the highest superconducting transition temperature reaches 119 K, exactly at the lowest pressure at which this structure is stable classically. Our results presented here show that quantum anharmonicity plays an important role also in \pstmmc\ \ScH\ by considerably increasing the H–H distance, which leads a large enhancement of the electron–phonon coupling and $T_c$. The paper is organized as follows: Sec. \ref{sec:methodolgy} describes the theoretical framework of our anharmonic {\it ab initio} calculations, Sec. \ref{sec:computational_details} overviews the computational details of our calculations, Sec. \ref{sec:results} presents the results of the calculations, and Sec. \ref{sec:conclusions} summarizes the main conclusions of this work.

\section{Methodology}
\label{sec:methodolgy}

The effect of quantum ionic fluctuations and anharmonicity is estimated using the stochastic self-consistent harmonic approximation (SSCHA) code\cite{monacelli2021stochastic}, whose theoretical basis was developed in Refs. \cite{34,35,51,52}. In this section we briefly review the SSCHA method  as well as the theoretical framework followed for estimating the superconducting critical temperature including anharmonicity.

\subsection{The stochastic self-consistent harmonic approximation}

The SSCHA is a quantum variational method that minimizes the 
free energy of the system calculated with a trial density matrix $\rhoschatrial$:
\begin{equation}
    \Fcal[\rhoschatrial] = \braket{K + V(\bR)}_{\rhoschatrial} - TS[\rhoschatrial].
    \label{eq:sscha_f}
\end{equation}
In the equation above, $K$ is the ionic kinetic energy, $T$ the temperature, and $S[\rhoschatrial]$ the entropy calculated with the trial
density matrix. The trial density matrix is parametrized with
centroid positions $\rschatrial$ and  auxiliary force constants $\phischatrial$. The former determine the average ionic positions and the latter are related to the broadening
of the ionic wave functions around $\rschatrial$. By minimizing $\Fcal[\rhoschatrial]$ with respect to $\rschatrial$ and $\phischatrial$, as well as calculating the stress tensor from $\Fcal[\rhoschatrial]$, the SSCHA code can optimize the crystal structure, including lattice degrees of freedom, fully including ionic quantum effects and anharmonicity at any target pressure.

Phonon frequencies within the SSCHA should not be calculated by diagonalizing the auxiliary force constants $\phischatrial$, but from the dynamical
extension of the theory\cite{51,monacelli2021timedependent,lihm2021gaussian}, which allows to determine phonon frequencies from the peaks of the one-phonon spectral function. In the static limit, the peaks coincide with the $\Omega_\mu(\bq)$ frequencies, 
where $\Omega^2_\mu(\bq)$ are the eigenvalues of the Fourier 
transform of the free energy Hessian matrix divided by the masses of the atoms:
\begin{equation}
\DF_{ab}= \frac{1}{\sqrt{M_aM_b}} 
\left[ \frac{\partial^2F}{\partial\Rcal^a\partial\Rcal^b}\right]_{\bRcaleq},
\label{eq:df}
\end{equation}
where $M_a$ is the mass of atom $a$.
In Eq. \eqref{eq:df} $F$ is assumed to be the free energy at the minimum, while $\bRcaleq$ the centroid positions that minimize Eq. \eqref{eq:sscha_f}.  If $\Omega_\mu(\bq)$ is imaginary the lattice is unstable in the quantum anharmonic energy landscape, as in that case the free energy is not a minimum along the lattice distortion determined by the corresponding eigenvector.

\subsection{Calculation of the superconducting transition temperature}

We evaluate $T_c$ with the Allen–Dynes\cite{53} modified McMillan equation,
\begin{equation}
T_c = \frac{f\textsubscript{1}f\textsubscript{2}\,\omega\textsubscript{log}}{1.2} \exp \left[ -\frac{1.04(1+\lambda)}{\lambda-\mu^*(1+0.62\lambda)} \right],
\label{eq1}
\end{equation}
where $\lambda$ is the electron-phonon coupling constant and $\mu^*$ effectively parametrizes the electron Coulomb repulsion\cite{54}. Despite its simplicity, this equation has led $T_c$ values in good agreement with experiments in hydrogen-rich compounds\cite{47}. $\lambda$ is calculated from the Eliashberg function $\alpha^{2}F(\omega)$ as
\begin{equation}
\lambda = 2 {\int_0^\infty d\omega \frac{\alpha^{2}F(\omega)}{\omega}}.
\label{eq2}
\end{equation}
Other parameters entering Eq. \eqref{eq1} are calculated as follows: 
\begin{eqnarray}
    \omega\textsubscript{log} & = & \exp \left( \frac{2}{\lambda} \int d\omega \frac{\alpha^2F(\omega)}{\omega} \log\omega \right), \\
    f_1 & = & \left[ 1 + (\lambda / \Lambda_1)^{3/2} \right]^{1/3}, \\
    f_2 & = & 1 + \frac{(\bar{\omega}_2/\omega\textsubscript{log} - 1) \lambda^2}{\lambda^2 + \Lambda_2^2}.
\end{eqnarray}
$\Lambda_1$, $\Lambda_2$, and $\bar{\omega}_2$ are given by
\begin{eqnarray}
    \Lambda_1 & = & 2.46 (1 + 3.8\mu^*) \\
    \Lambda_2 & = & 1.82 (1 + 6.3\mu^*)(\bar{\omega}_2/\omega\textsubscript{log}) \\
    \bar{\omega}_2 & = & \left[ \frac{2}{\lambda} \int d\omega \alpha^2F(\omega) \omega \right]^{1/2}.
\end{eqnarray}

The Eliashberg function is calculated as
\begin{eqnarray}
    && \alpha^{2}F(\omega) = \frac{1}{2 N(0) N_q N_k} \sum_{\substack{\mu \bq \\ {\bk}nm \\ \bar{a}\bar{b}}}
    \frac{\epsilon_{\mu}^{\bar{a}}(\bq) \epsilon_{\mu}^{\bar{b}}(\bq)^*}{\omega_{\mu}(\bq) \sqrt{M_{\bar{a}}M_{\bar{b}}}} \nonumber \\
    && \times
    d^{\bar{a}}_{{\bk}n,{\bk}+{\bq}m} d^{\bar{b}*}_{{\bk}n,{\bk}+{\bq}m} \delta(\varepsilon_{{\bk}n})
   \delta(\varepsilon_{{\bk+\bq}m}) \delta(\omega -  \omega_{\mu} (\bq)). \nonumber \\
\label{eq:eliashberg}
\end{eqnarray}
In the equation above  $d^{\bar{a}}_{{\bk}n,{\bk}+{\bq}m} = \bra{{\bk}n}
\delta V_{KS} / \delta R^{\bar{a}}(\bq) \ket{{\bk}+{\bq}m}$, where 
$\ket{{\bk}n}$ is a Kohn-Sham state with energy $\varepsilon_{{\bk}n}$
measured from the Fermi level, $V_{KS}$ is Kohn-Sham potential, and
$R^{\bar{a}}(\bq)$ is the Fourier transformed displacement of atom
$\bar{a}$; $N_k$ and $N_q$ are the number of electron
and phonon momentum points used for the BZ sampling; $N(0)$ is the density of states 
at the Fermi level; and  $\omega_{\mu} (\bq)$ and 
$\epsilon_{\mu}^{\bar{a}}(\bq)$ represent phonon frequencies and polarization vectors.
The combined atom and Cartesian indexes with a bar ($\bar{a}$) 
only run for atoms inside the unit cell.
In this work, the Eliashberg function is calculated both at the harmonic or anharmonic levels, respectively, by plugging into Eq. \eqref{eq:eliashberg} the harmonic phonon frequencies and polarization vectors or their anharmonic counterparts obtained diagonalizing $\bDF$. It should be noted as well that the derivatives of the Kohn-Sham potential entering the electron-phonon matrix elements are calculated at different positions in the classical harmonic and quantum anharmonic calculations: in the former they are taken at the $\bR_0$ positions that minimize  $V(\bR)$, while in the latter at the $\bRcaleq$ positions that minimize instead $\Fcal[\rhoschatrial]$. 

\section{Computational details}
\label{sec:computational_details}

All first-principles calculations were performed within DFT\cite{45} and the generalized gradient approximation (GGA) as parametrized by Perdew, Burke, and Ernzerhof (PBE)\cite{46,47}. Harmonic phonon frequencies were calculated making use of density functional perturbation theory (DFPT)\cite{DFPT_S.Baroni} as implemented in the Quantum ESPRESSO code\cite{48}. We used PAW pseudopotentials\cite{49,50}, including $11$ electrons in the valence for Sc. The SSCHA\cite{monacelli2021stochastic} minimization requires the calculation of energies, forces, and stress tensors in supercells. These were calculated as well within DFT at the PBE level with Quantum ESPRESSO, making use of the same pseudopotentials. The plane-wave basis cutoff was set to 80 Ry for the kinetic energy and to 800 Ry for the density. Brillouin zone (BZ) integrations were performed on a 21$\times$21$\times$14 Monkhorst-Pack $\bk$-point mesh\cite{53}, using a smearing parameter of 0.01 Ry, for the unit cell harmonic phonon calculations. The SSCHA calculations were performed using a 2$\times$2$\times$1 supercell containing 56 atoms at 0 K, yielding dynamical matrices on a commensurate 2$\times$2$\times$1 $\bq$-point grid. A 50 Ry energy cutoff and a 6$\times$6$\times$6 k-point mesh for the Brillouin zone (BZ) integrations were sufficient in the supercell to converge the SSCHA gradient. Harmonic phonon frequencies and electron–phonon matrix elements were calculated on a grid of 6$\times$6$\times$4 points. The difference between the harmonic and anharmonic dynamical matrices in the 2$\times$2$\times$1 phonon-momentum grid was interpolated to a 6$\times$6$\times$4 grid. Adding the harmonic 6$\times$6$\times$4 grid dynamical matrices to the result, the anharmonic 6$\times$6$\times$4 $\bq$-grid dynamical matrices were obtained. The Eliashberg function in Eq. \eqref{eq:eliashberg} was calculated with a finer 30$\times$30$\times$30 $\bk$-point grid, with a Gaussian smearing of 0.008 Ry for the electronic Dirac deltas. 

\section{Results and discussion}
\label{sec:results}

\begin{figure}
\includegraphics[scale=0.6]{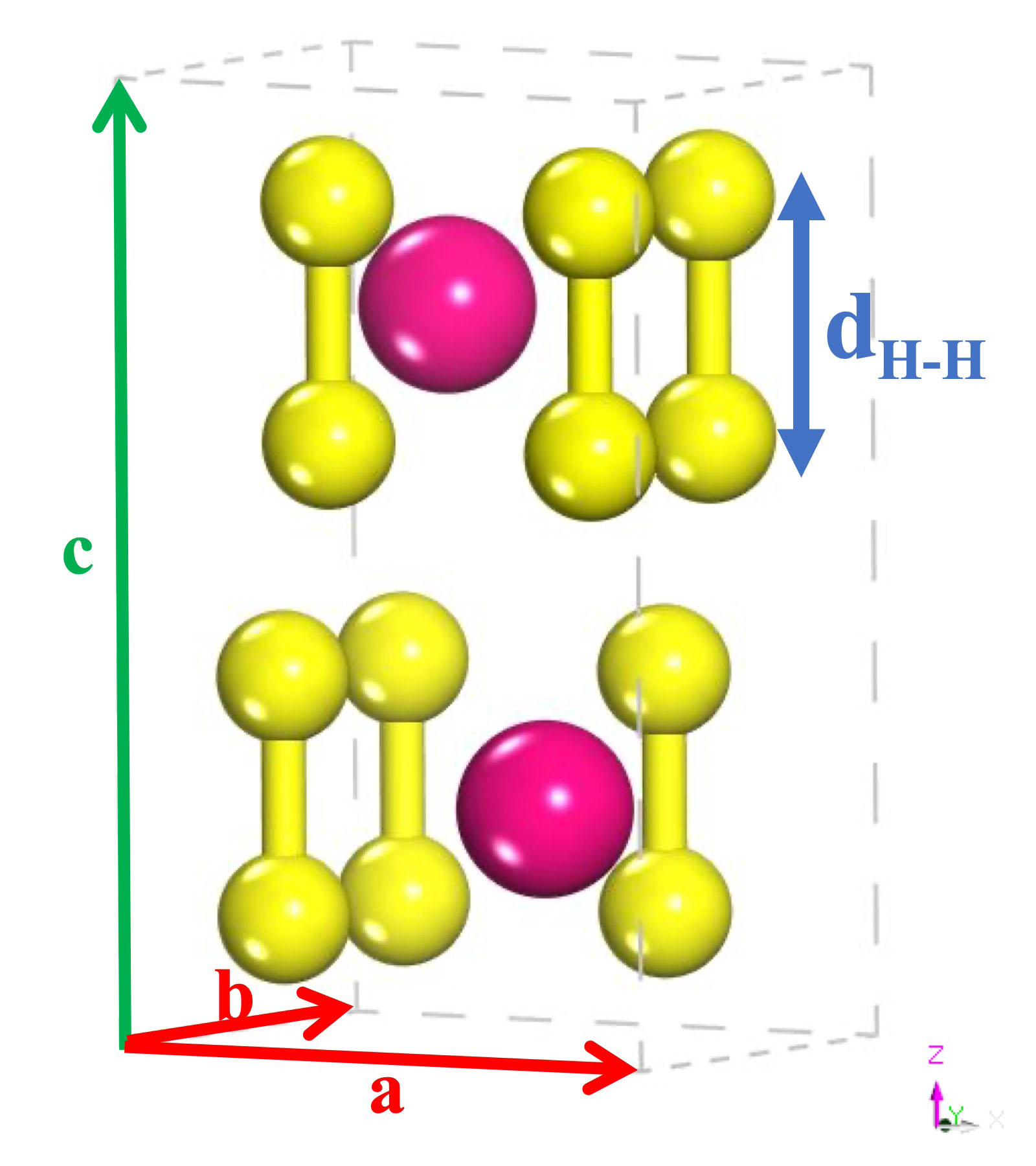}
\caption{(Color online) Crystal structure of \pstmmc\ \ScH\ in its hexagonal lattice. Large pink balls are Sc atoms, and small golden balls are H atoms. The distance of hydrogen pairs $d_{H-H}$ and lattice parameters $a$, $b$, and $c$ are marked.}
\label{1.jpg} 
\end{figure}

As mentioned above, in the classical harmonic approximation, the atomic positions are those determined by the minimum of $V(\bR)$. If quantum anharmonic effects are considered instead, the atomic positions, i.e. the Wyckoff positions, are determined by the minimum of the energy that includes the vibrational contribution,  the zero-temperature limit of the free energy in Eq. \eqref{eq:sscha_f}. The hexagonal crystal structure of \pstmmc\ \ScH\ is shown in Fig. \ref{1.jpg}. Sc atoms occupy the $2d$ Wyckoff positions, completely fixed by symmetry, and H atoms the $12k$ sites, a representative of which can be written as $(x,2x,z)$. Thus, in total there are four free parameters in the crystal structure, the $a$ and $c$ lattice parameters of the hexagonal cell, plus the $x$ and $z$ free parameters in the Wyckoff positions occupied by the H atoms. As we performed the SSCHA minimization keeping symmetries, quantum anharmonic effects optimize precisely these four free parameters. Our results, as summarized in Table \ref{tab:my-table}, show that quantum anharmonic effects significantly modify the crystal structure of \pstmmc\ \ScH. The most important effect is that the H–H distance between the hydrogen atoms forming the H$_2$ units, $d_{H-H}$, is largely increased by quantum anharmonicity. At 130 GPa $d_{H-H}$ is increased by a non-negligible 5.5\%. It is also noteworthy that quantum anharmonic effects stretch the out-of-plane $c$ parameter more than the in-plane $a$ parameter. At all pressures studied, quantum anharmonic effects impose a similar change in the structure.


\begin{table}
\caption{Calculated structural parameters of the \pstmmc\ \ScH\ within the classical harmonic approximation and at the quantum anharmonic level at 130 GPa. We give the lattice parameters $a$ and $c$, as well as the $x$ and $z$ parameters that determine the positions of the H atoms in the $12k$ Wyckoff sites.}
\label{tab:my-table}
\centering
\begin{tabular}{cccccc}
\hline
\hline
           & $a=b$ (\AA) & $c$ (\AA)  & $x$     & $z$     &  $d_{H-H}$ (\AA)     \\ 
\hline 
Harmonic   & 3.429  & 4.284 & 0.166 & 0.369 & 1.022 \\   
Anharmonic & 3.450  & 4.328 & 0.165 & 0.375 & 1.078 \\
\hline
\hline
\end{tabular}
\end{table}

\begin{figure*}
\includegraphics[scale=1.15]{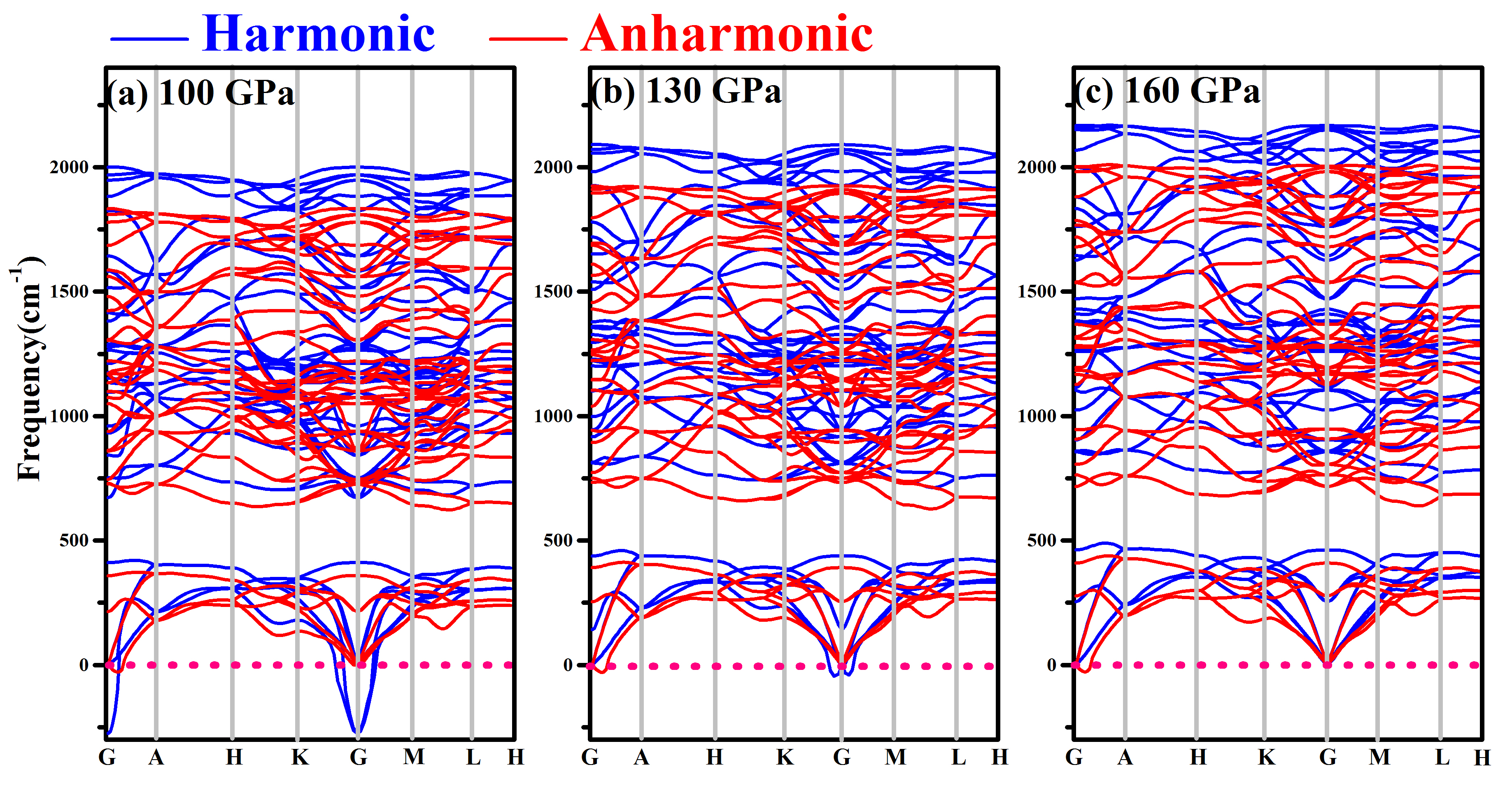}
\caption{(Color online) Phonon spectra of \pstmmc\ \ScH\ at different pressures both at the classical harmonic and quantum anharmonic levels. The area under the red dotted line marks the region with imaginary phonon frequencies, which are depicted as negative frequencies. }
\label{2.jpg} 
\end{figure*}

Quantum anharmonic effects not only affect the structure, also the phonon spectra. Fig. \ref{2.jpg} compares the harmonic phonons obtained diagonalizing the harmonic dynamical matrix, 
\begin{equation}
\DHA_{ab} = \frac{1}{\sqrt{M_a M_b}} 
\left[ \frac{\partial^2V(\bR)}{\partial R^a \partial R^b} \right]_{\bR_0},
\label{eq:df_h}
\end{equation}
with the anharmonic ones obtained diagonalizing instead Eq. \eqref{eq:df}. The anharmonic correction of the phonon spectra given by the SSCHA is huge. The SSCHA renormalization of the phonons clearly indicates that almost every mode is softened by anharmonicity. The softening of the high-energy phonons of mainly H character, separated with a gap from the low-energy modes of mostly Sc character, is consistent with the stretching of the H$_2$ units and the weakening of the H-H bonds. 

Interestingly, below 130 GPa, the harmonic approximation predicts that a phonon mode with $E_{2g}$ symmetry at the $\Gamma$ point, which both affects Sc and H displacements, becomes unstable. This instability completely disappears in the SSCHA calculation. The small instability that appears at the $\Gamma$ point in the anharmonic case for the whole range we studied and the harmonic case at 130 GPa are artifacts of the Fourier interpolation. We thus can say that quantum anharmonic effects stabilize this structure below 130 GPa. Our further anharmonic calculations show that this structure remains dynamically stable down to at least 85 GPa, which means we can get this structure 45 GPa lower than expected with   classical harmonic calculations. Our interpolated quantum anharmonic phonon spectra develops an instability below 85 GPa between high symmetry points H and K in the BZ. Whether this instability is an artifact of the Fourier interpoation or it is a real instability may be discerned performing SSCHA calculations on larger supercells.


Once the renormalized anharmonic phonon spectra has been obtained using the SSCHA, anharmonic effects can be easily incorporated into the electron-phonon coupling calculations as explained in Sec. \ref{sec:methodolgy}. We calculate $T_c$ with typical values of $\mu^*$ such as 0.10 and 0.15, and obtain high critical temperatures of around 100 K. For a given value of $\mu^*$, in the anharmonic calculation $T_c$ approaches a linear decrease with increasing pressure in the 120-140 pressure range, while it remains approximately constant in the 100–120 GPa and 140-160 GPa pressure ranges. On the contrary, there is a clear decreasing trend in the 130-160 GPa range at the harmonic level, the same behavior described in previous works\cite{29,30,31}. Due to the instabilities present in the harmonic phonon spectra below 130 GPa, $T_c$ cannot be calculated below this pressure in the harmonic case. The fact that $T_c$ soars with pressure lowering is in accordance with the general goal of obtaining materials with high critical temperature at lower pressures. The electron-phonon coupling constant $\lambda$ evolves under pressure practically as $T_c$, which suggests that the evolution of $T_c$ is governed by $\lambda$. 

\begin{figure}
\includegraphics[scale=1.1]{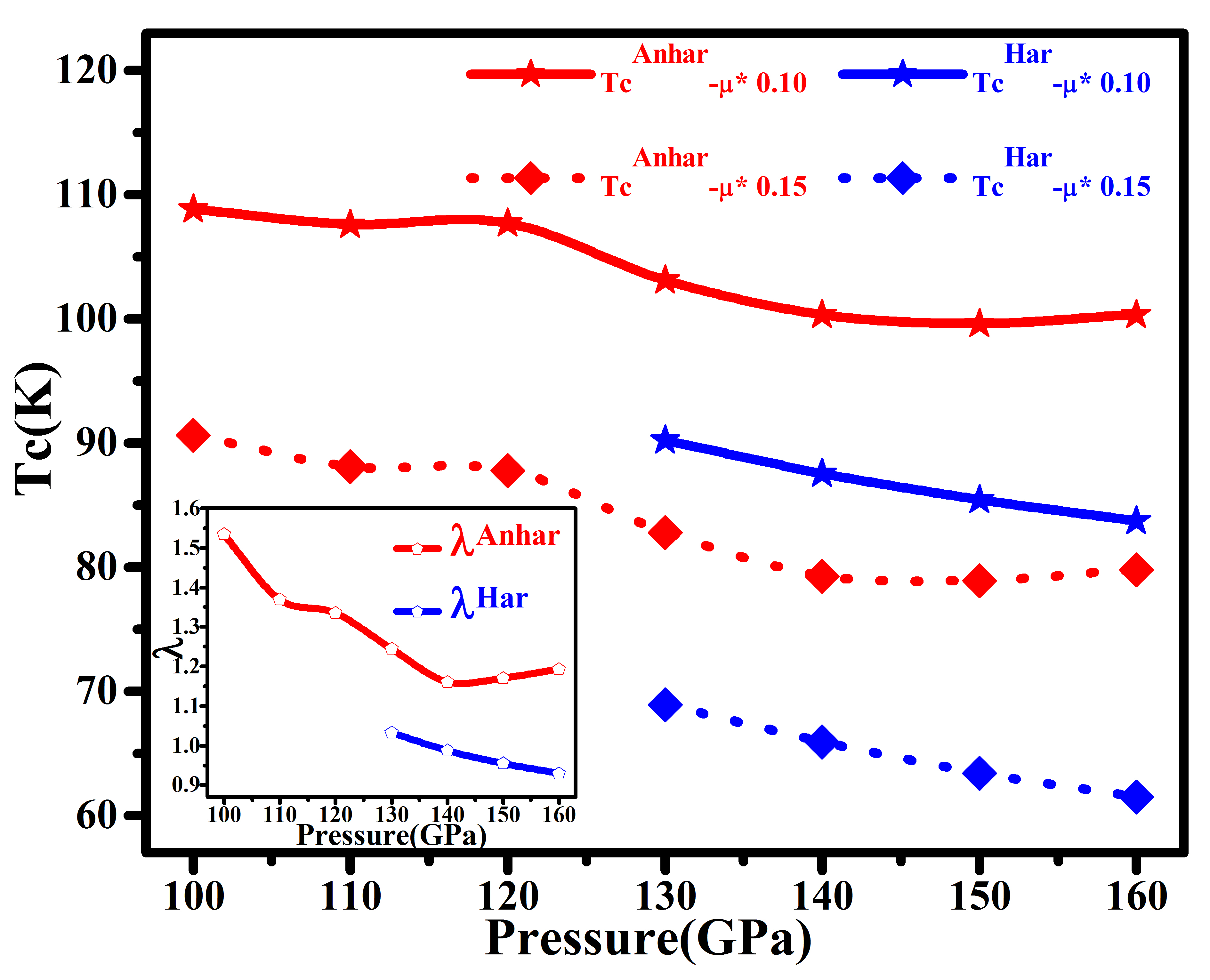}
\caption{(Color online) Superconducting critical temperature $T_c$ and electron-phonon coupling constant $\lambda$ as a function of pressure at the harmonic and anharmonic levels.}
\label{3.jpg} 
\end{figure}

The superconducting behavior of \pstmmc\ \ScH\ changes radically when the anharmonic renormalization of the Wyckoff positions and phonons is considered. As shown in Fig. \ref{3.jpg}, anharmonicity makes the electron-phonon coupling constant reach a value of 1.53 at 100 GPa and 1.19 at 160 GPa, a 28\% larger than the harmonic value in the latter case. This supposes a huge anharmonic enhancement of the electron–phonon coupling that brings $T_c$ about 16(18) K higher than the harmonic value for $\mu^*=0.10(0.15)$ at 160 GPa. A similar correction is observer at other pressures. Remarkably, this large anharmonic enhancement of superconductivity is exactly the opposite effect to the one described in other superconducting hydrides\cite{hou2021strong,33,34,35}.

\begin{figure}
\includegraphics[scale=1]{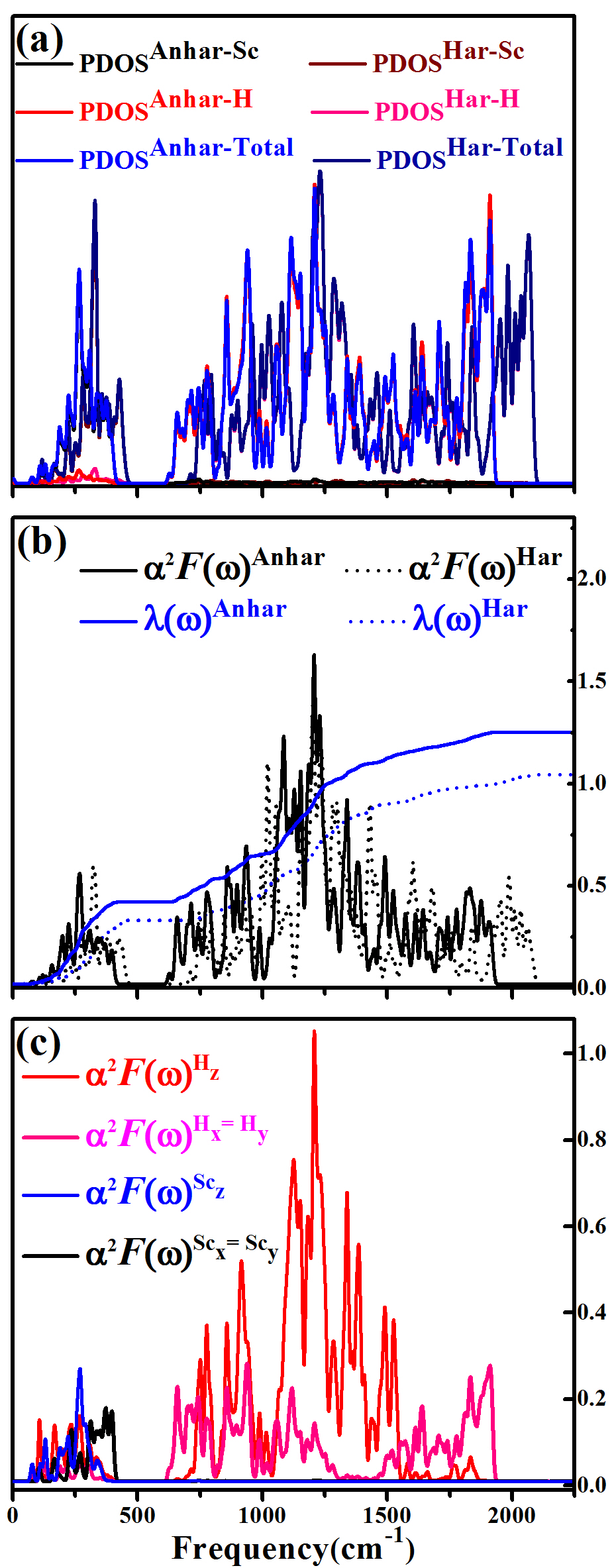}
\caption{(Color online) Superconducting properties of \pstmmc\ \ScH. (a) Total phonon density of states (PDOS), and the contribution from H and Sc atoms (black solid lines) at 130 GPa. For comparison, the harmonic results are also shown. (b) Anharmonic (solid line) and harmonic (dotted line) Eliashberg spectral function $\alpha^{2}F(\omega)$ (black line) and its integral $\lambda(\omega)$ (blue line) at 130 GPa. (c) The projection of $\alpha^{2}F(\omega)$ at the anharmonic level at 130 GPa into H and Sc atoms along $x$, $y$, and $z$ Cartesian directions.}
\label{4.jpg} 
\end{figure}

In order to deepen into the origin of such anharmonic enhancement of superconductivity, it is convenient to have a look into the projected phonon density of states (PDOS), $\alpha^{2}F(\omega)$, and the 
\begin{equation}
    \lambda(\omega) = 2 {\int_0^\omega d\Omega \frac{\alpha^{2}F(\Omega)}{\Omega}}
\end{equation}
integrated electron-phonon coupling constant (see Fig. \ref{4.jpg}). The contribution to the PDOS from H and Sc atoms is clearly separated. All the contribution in the high-frequency region comes from H atoms, while the low-frequency vibrations (below 500 $cm^{-1}$) are dominated by the heavy Sc atoms. As expected in this type of superconductors, H atoms have the largest contribution to $\lambda$, around 70\% of the total contribution. As shown in Fig. \ref{4.jpg}(b), the electron-phonon coupling constant is enhanced in the anharmonic case mostly because of the phonon softening induced by anharmonicity, which redshifts the Eliashberg function. It is interesting to project further the Eliashberg function into Cartesian directions. As evidenced in Fig. \ref{4.jpg}(c), the projection of H atoms into the $z$ direction has the largest contribution to the electron-phonon coupling constant. This suggests that the anharmonic enhancement of superconductivity in \pstmmc\ \ScH\ is driven by the anharmonic stretching of the H$_2$ molecular-like units, which occurs in the $z$ direction (see Fig. \ref{1.jpg}).



\begin{figure}
\includegraphics[scale=1]{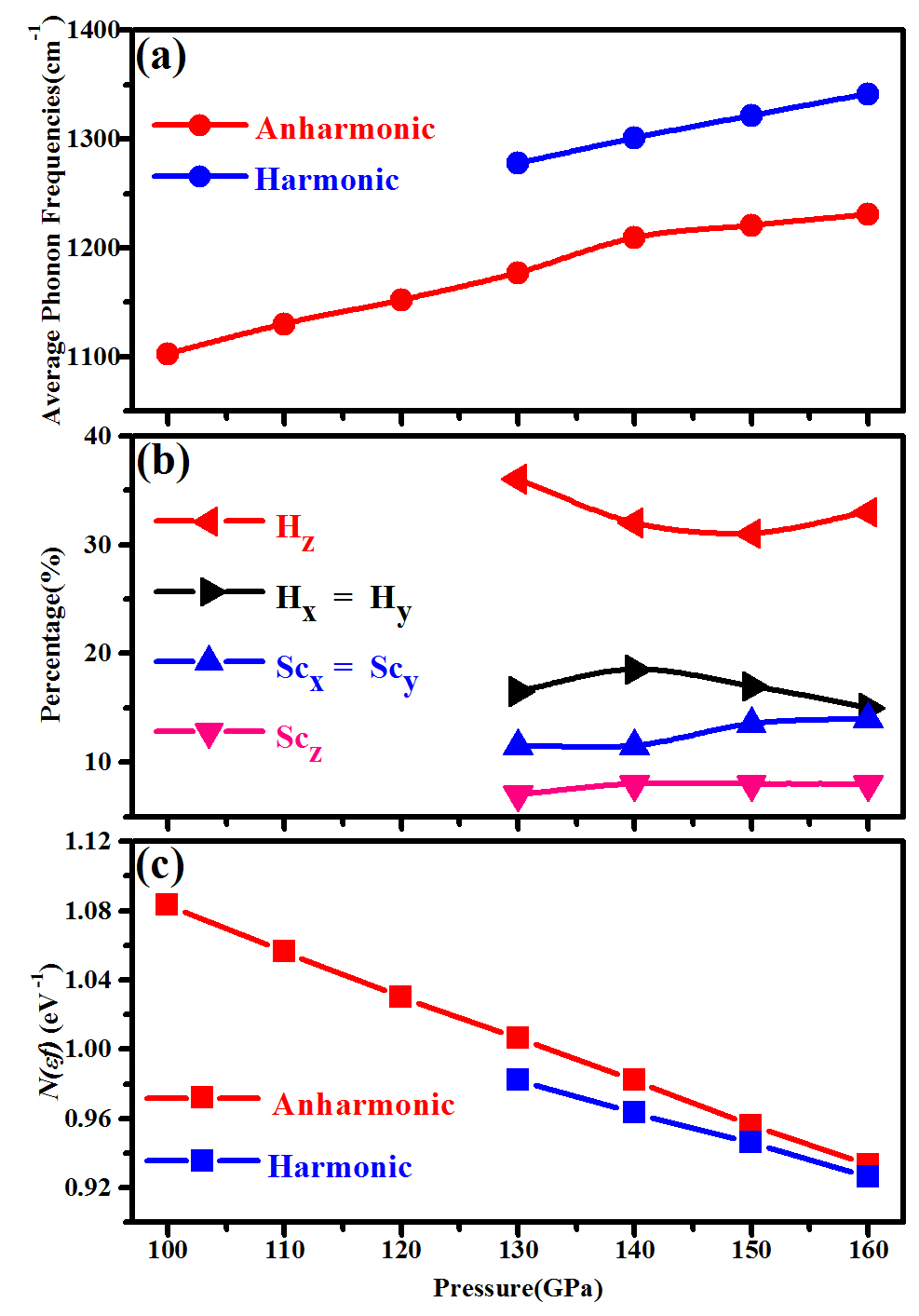}
\caption{(Color online) (a) Calculated average phonon frequency $\left\langle\omega\right\rangle$ both at the harmonic and anharmonic level. (b) The percentages of contributions from H and Sc atoms along $x$, $y$, and $z$ Cartesian directions to the increase of the electron-phonon coupling constant due to anharmonicity. (c) The DOS at the Fermi level at the harmonic and anharmonic level.}
\label{5.jpg} 
\end{figure}

This conclusion is further supported by analyzing in more detail the structure of $\lambda$. The electron–phonon coupling constant, indeed, can be written as
\begin{equation}
\lambda = \frac{N(0)D^2}{M\left\langle\omega\right\rangle^2},
\label{eq7}
\end{equation}
where $N(0)$ is the density of states (DOS) at the Fermi level, $D$ is the so-called deformation potential that measures in an averaged way the strength of the electron-phonon matrix elements, $M$ is the atomic mass, and $\omega$ is an effective average phonon frequency. The average phonon frequency, calculated as
\begin{equation}
\left\langle\omega\right\rangle = \frac{\int_0^\infty g(\omega) \omega d\omega}{\int_0^\infty g(\omega) d\omega},
\label{eq8}
\end{equation}
where $g(\omega)$ is the PDOS, reflects the phonon softening induced by anharmonicity. Our results give $(\left\langle\omega\right\rangle\textsubscript{harmonic} /\left\langle\omega\right\rangle\textsubscript{anharmonic})^2$ = 1.18 at 130 GPa. This ratio is similar at other pressures. As shown in Fig. \ref{5.jpg} (c) it is remarkable that the DOS at the Fermi level is always larger for the quantum (anharmonic) structure than for the classical (harmonic) structure. This is in line with the conclusion drawn recently\cite{44}, which suggests that nuclear quantum effects can make stable crystal structures with larger DOS at the Fermi level, enhancing thus the electron-phonon coupling constant. The phonon softening and the slight increase in $N(0)$ explain mainly the increase in the anharmonic $\lambda$, without requiring a large modification of the electron-phonon matrix elements themselves. This was not the case, for instance, in the metallic and molecular $Cmca-4$ phase of hydrogen, where quantum effects in the atomic positions induced a remarkable change in the electronic band structure clearly affecting the deformation potential\cite{41}.


By calculating the contribution to $\lambda$ with the projected Eliashberg functions, we can determine explicitly from which atoms the increase of the electron-phonon coupling constant is coming from once quantum anharmonic effects are considered. As shown in Fig. \ref{5.jpg}(b), more than a 30\% of the total increase of $\lambda$ comes from the enhanced contribution of the H atoms along the $z$ direction, clearly the largest contribution to the enhancement of $\lambda$. This confirms that the stretching of the H$_2$ units is the key in the quantum anharmonic enhancement of superconductivity.



\section{Conclusions}
\label{sec:conclusions}

In summary, this work demonstrates the importance of quantum effects and anharmonicity on the structural and superconducting properties of \pstmmc\ \ScH\ under high pressure. Quantum anharmonic effects strongly affect the position of hydrogen atoms in the crystal by considerably stretching the H$_2$ molecular-like units by a non-negligible 5\%. Consequently, the phonon spectra are strongly affected, with a pronounced softening in the majority of the modes. However, quantum anharmonic effects make the structure dynamical stable down to 85 GPa, which is about 45 GPa lower than expected classically. Moreover, anharmonicity increases the electron-phonon coupling constant by no less than 18\% and $T_c$ by at least 15\% in the range of 130–160 GPa. We determine that the increase is directly a consequence of the stretching of the H$_2$ units. Our results suggest that in hydrogen-based superconductors that have free parameters in the crystal structure, quantum anharmonic effects can induce a large structural modification that yields a general softening of the phonon spectra as well as an enhanced DOS at the Fermi level, which both contribute to enhance the superconducting critical temperature.

\section*{Acknowledgements}
This research was supported by the European Research Council (ERC) 
under the European Unions Horizon 2020 research and innovation programme (grant agreement No. 802533).

\section*{Data Availability Statement}
The data that support the findings of this study are available from the corresponding author upon reasonable request.

%


\end{document}